# TCAD Modeling of Cryogenic nMOSFET ON-State Current and Subthreshold Slope

Prabjot Dhillon
*Electrical Engineering*
*San Jose State University*
San Jose, USA
prabjot.dhillon@sjsu.edu

Nguyen Cong Dao
*Electrical and Info. Engineering*
*The University of Sydney*
Sydney, Australia
nguyen.dao@sydney.edu.au

Philip H. W. Leong
*Electrical and Info. Engineering*
*The University of Sydney*
Sydney, Australia
philip.leong@sydney.edu.au

Hiu Yung Wong*
*Electrical Engineering*
*San Jose State University*
San Jose, USA
hiuyung.wong@sjsu.edu

*Abstract*— In this paper, through careful calibration, we demonstrate the possibility of using a single set of models and parameters to model the ON current and Sub-threshold Slope (SS) of an nMOSFET at 300K and 5K using Technology Computer-Aided Design (TCAD). The device used is a 0.35μm technology nMOSFET with W/L=10μm/10μm. We show that it is possible to model the abnormal SS by using interface acceptor traps with a density less than $2\times10^{12}cm^{-2}$. We also propose trap distribution profiles in the energy space that can be used to reproduce other observed SS from 4K to 300K. Although this work does not prove or disprove any possible origin of the abnormal SS, it shows that one cannot completely rule out the interfacial traps as the origin and it shows that interfacial traps can be used to model the abnormal SS before the origin is fully understood. We also show that Drain-Induced-Barrier-Lowering (DIBL) is much reduced at cryogenic temperature due to the abnormal slope and the device optimization strategy might need to be revised.

## I. Introduction

Cryogenic CMOS is the most promising technology to enable effective and large-scale quantum computing [1][2] and deep space exploration [3]. Most of the cryogenic device studies rely on analytical modeling [4][5] because TCAD modeling of cryogenic CMOS is still not mature due to a lack of well-calibrated parameters at cryogenic temperatures [5], unknown physics (e.g. at cryogenic temperature, sub-threshold slope, SS, does not scale with temperature [6][7] and it is dubbed as "abnormal SS" in this paper), and convergence difficulties [8].

Due to convergence issues, comprehensive TCAD simulation is usually only performed down to 77K [9]. With simplified models, it is still very difficult to reach liquid-He temperature (e.g. 15K in [10]), although a more advanced model for field-dependent incomplete ionization has been demonstrated in a 3D resistor simulation at 4.2K [11]. Therefore, it is very desirable to demonstrate the possibility of modeling realistic MOSFET at liquid-He temperature.

SS of a MOSFET is expected to scale with temperature (SS ~ nkT/q, where n, k, T, and q are the ideality factor, Boltzmann constant, temperature, and elementary charge, respectively). Therefore, at cryogenic temperature, it is expected to have a very steep SS (e.g. 0.8mV/dec at 4K for n=1) and thus improve the $I_{ON}/I_{OFF}$ trade-off. However, various experimental results from different groups and technologies show that the SS is saturated at about 10mV/dec to 15mV/dec, which corresponds to the SS at 50K to 70K range, as the temperature decreases. There have been two major theories to explain the abnormal SS. The first one, which is more promising and dominating, attributes the phenomenon to the existence of Si band tail [6][7]. The second one attributes the SS degradation to the traps at the oxide/channel interface [12] and this was also supported by its correlation with the increase in 1/f noise at cryogenic temperatures [13]. However, due to the convergence difficulties in TCAD, such theory has been only studied analytically and a very large interface trap density is required to explain the degradation.

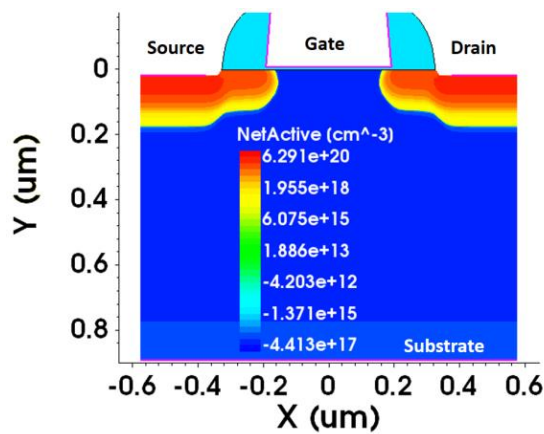

Figure 1: Cross-section of the simulated 0.35μm technology nMOSFET. $L_G$=0.35μm is shown for clarity. $L_G$=10μm is used in the calibration (Fig. 2).

Therefore, in this work, we try to use TCAD to model nMOSFET transistor ON-state current and also SS with proper numerical and model parameter settings. We explore the possibility of using interface traps to model the abnormal subthreshold slope and its applicability in different regimes which is not impossible with analytical solutions.

## II. TCAD Modeling of Experimental Data

TCAD Sentaurus is used in this work [14]. We measured the $I_D$-$V_G$ curves of a 0.35μm technology nMOSFET with W/L=10μm/10μm at 300K and 5K with $V_D$ = 0.1V [15]. Although the fabrication conditions are unknown, since 0.35μm technology is a mature and standard process, modifications are made based on [16] to obtain reasonable matching. Fig. 1 shows the device simulated using SProcess (note that for clarity, $L_G$=0.35μm is shown but $L_G$=10μm is used in the calibration). The structure is realistic such as with oxide thickening at gate edge due to Poly Reox and with source/drain recess due to spacer overetch. The gate oxide thickness, $t_{ox}$, is 7.6nm.

---
*Corresponding Author: hiuyung.wong@sjsu.edu

SDevice is used to simulate the $I_D$-$V_G$ curves. For good convergence, transient simulation with extrapolation and Backward-Euler is used. To improve numerical stability, 80-bit extended precision is used with the minimum carrier density solution set to $10^{-2000}$cm$^{-3}$ in the Newton iteration. An iterative linear solver is used. Trap level discretization is increased to 1000 in the energy space for accurate trap modeling.

Fermi statistics is turned on. Philip Unified Mobility model (PhuMob) and Lombardi Model for surface scattering are used for mobility calculations. PhuMob parameters calibrated in [11] are used. Since channel boron doping is expected to be fully ionized in *ON*-state and the LDD doping is in the order of $10^{20}$cm$^{-3}$ near the surface, the incomplete ionization model is not turned on. Moreover, since only $V_D$ = 0.1V is used with L = 10μm, the velocity saturation model is not turned on.

Fig. 2 shows that the *ON*-state current and SS can be matched well at 300K and 5K. Two major calibrations were performed. Firstly, the enhanced Lombardi model's acoustic phonon scattering part parameters are adjusted [14]. The acoustic phonon scattering part has the following formula,

$$\mu_{ac} = \frac{B}{F_\perp} + \frac{C\big((N+N_2)/N_0\big)^\lambda}{F_\perp^{1/3}(T/300K)^k} \quad (1)$$

where $N$ is the total doping concentration, $F_\perp$ is the normal component of the electric field at the interface, $T$ is the temperature, and B, C, $\lambda$, $k$, $N_2$, and $N_0$ are fitting parameters. The default values are used except that the *C* parameter is reduced from 580cm$^{5/3}$V$^{-2/3}$s$^{-1}$ to 340cm$^{5/3}$V$^{-2/3}$s$^{-1}$ to match the 300K data and the *k* parameter is reduced from 1 to 0.45 to reduce the temperature dependency in order to fit both the 300K and 5K data.

Although the ON-state current matches well, the simulated SS (~1mV/dec) and threshold voltage are too small at 5K as shown in Fig. 2 (dash lines). Therefore, secondly 6×10$^{13}$cm$^{-2}$eV$^{-1}$ acceptor traps are assigned uniformly between $E_C$-5meV and $E_C$+25meV at the oxide/silicon interface, where $E_C$ is the conduction band edge. This is equivalent to $1.8\times10^{12}$cm$^{-2}$ which is a reasonable value. The choice of the energy range and density of the trap is based on manual fitting, which will be explored more in the next section.

### III. INTERFACE TRAP EFFECTS ON SS

While interface acceptor traps can help model the abnormal SS at 5K, it is important to understand if the same set of traps can be used to model the SS at various temperatures. It is also important to investigate how the SS varies with oxide thickness, gate length, and drain voltage. It is worth noting that since the transient simulation is used, the trap capturing and emission times are taken into account properly in this simulation. Although the capturing cross-section is not calibrated at cryogenic temperature, the capture rate depends on temperature through thermal velocity:

$$c = \sigma v_{th,0}\sqrt{\frac{T}{300K}}n \quad (2)$$

where $\sigma$, $v_{th,0}$, $T$, and $n$ are the cross-section, thermal velocity at 300K, temperature, and carrier concentration, respectively.

#### A. Temperature Variation

A device of W/L = 1μm/5μm simulated in a simplified process flow is constructed to study the effect of traps on SS. $I_D$-$V_G$ curves are simulated with $V_D$=1mV. Two different oxide thicknesses were studied ($t_{ox}$ = 10nm and $t_{ox}$ = 2nm).

To fit the SS of various temperatures using traps, the highest temperature with abnormal SS is fitted first. The Femi-level locations at the start and end of the subthreshold region are first identified and a uniform trap (in the energy space) is assigned between these two locations. Then the next lower temperature curve is fitted similarly with the traps from the previous fitting(s) kept. This is repeated until the 4K curve. After that, the overall profile is further adjusted to obtain a smooth curve.

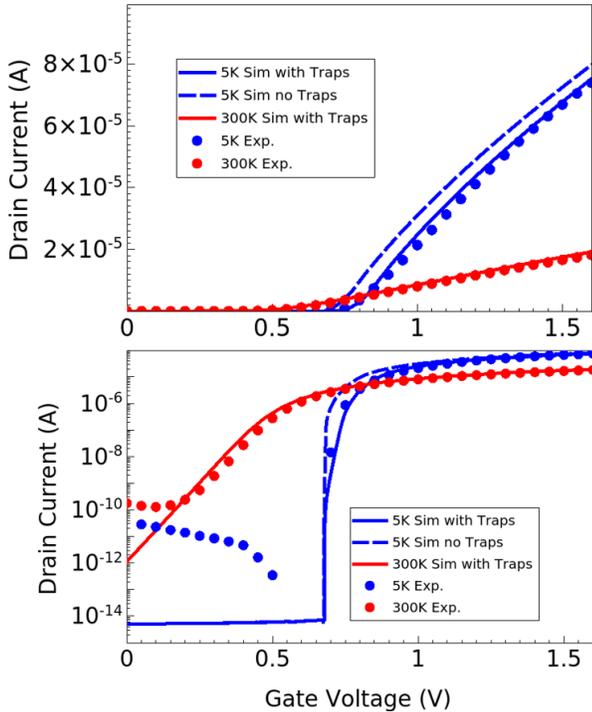

Figure 2: Experimental and simulated $I_D$-$V_G$ curves in linear (top) and logarithmic (bottom) scales of W/L=10μm/10μm at 300K and 5K. 6×10$^{13}$cm$^{-2}$eV$^{-1}$ SiO$_2$/Si interface acceptor traps is assigned uniformly between $E_C$-5meV and $E_C$+25meV in the "with trap" case.

Fig. 3 shows the $I_D$-$V_G$ curves with and without traps for $t_{ox}$ = 10nm. An optimized trap profile to reproduce the abnormal SS at various temperatures is deduced and shown in Fig. 4. The SS is extracted from the slope of the $I_D$-$V_G$ curves between $I_D$ = 10$^{-13}$ A and 10$^{-10}$A and plotted in Fig. 5 against the experimental values by Beckers et at. [6]. Note that the device used in [6] is of 28nm technology and the oxide thickness and substrate doping are unknown to us. Therefore, the purpose of the fitting is only to show the possibility of fitting SS by using one single set of trap distribution.

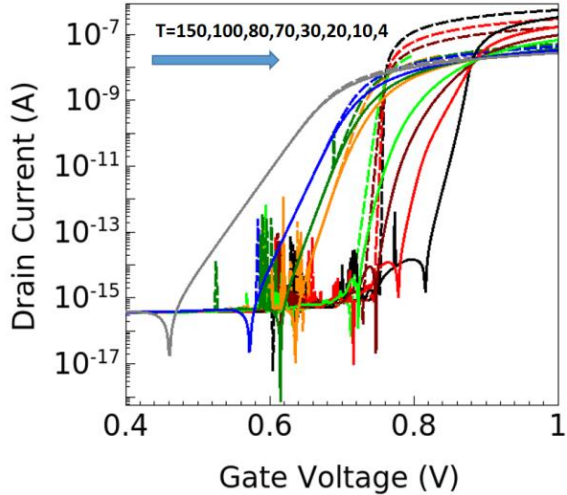

Figure 3: Simulations of transistor ($t_{ox}$=10nm) $I_D$-$V_G$ with $V_D$=1mV at various temperatures with (solid) and without (dash) traps. Single set of traps is used for all temperatures. For clarity, T>150K are not shown. This transistor is different from that in Fig. 1.

The device used in this study has a similar SS at 300K as that in [6]. By using the trap profiles showed in Fig. 4, the SS trend can be reproduced well. The total acceptor trap is integrated to be $2.47 \times 10^{11}$cm$^{-2}$, which is reasonably low.

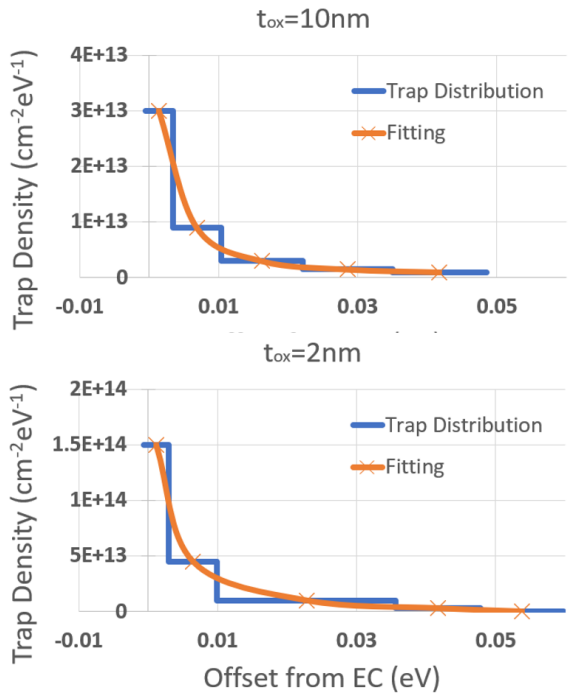

Figure 4: Trap distributions required to reproduce the abnormal SS at all temperatures for two different oxide thicknesses.

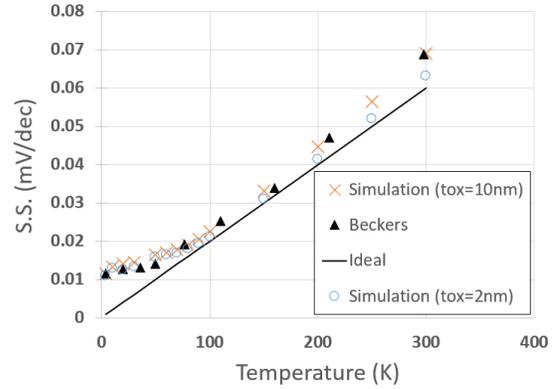

Figure 5: SS as a function of temperature. "Beckers" is from the experimental data in [6].

### B. Gate Insulator Variation

However, it is known that the effect of the acceptor trap on $I_D$-$V_G$ curves depends inversely on the gate oxide capacitance. For advanced technology such as that in [6], it is expected that more traps are needed to achieve the same effect. Therefore, $t_{ox}$ = 2nm is used and another optimal trap profile is deduced in Fig. 4 and the SS is plotted in Fig. 5. While the abnormal SS can be matched again (except at high temperature because of the better gate control in the simulation than in the experiment), the total integrated charge is $1.14 \times 10^{12}$cm$^{-2}$, which is about 5 times higher than the first profile. This agrees with the fact that the oxide capacitance is 5 times higher.

### C. Drain Voltage Variation

To study the effect of drain induced barrier lowering (DIBL) on the validity of using interface traps to model abnormal SS, a device with L=0.5μm, W=1μm, and $t_{ox}$=2nm by incorporating the trap profile from Fig. 4 ($t_{ox}$=2nm case) is simulated at various $V_D$. Fig. 6 shows the extracted SS as a function of temperature.

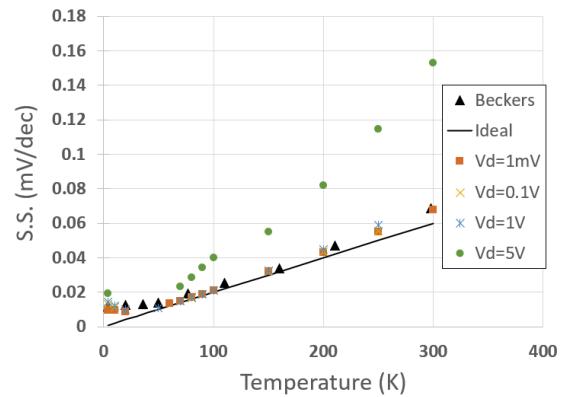

Figure 6: SS as a function of temperature at various $V_D$ for L=0.5μm and $t_{ox}$=2nm. "Beckers" is from the experimental data in [6].

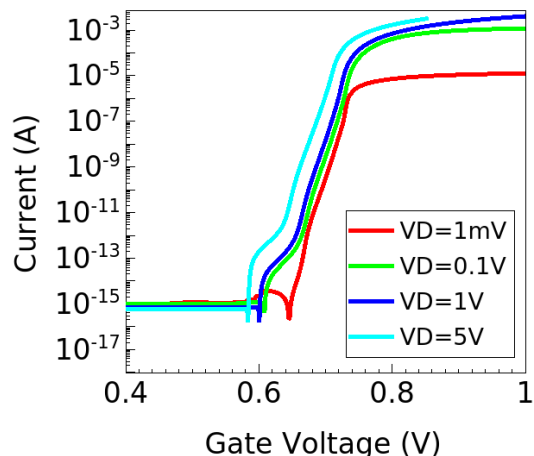

Figure 7: $I_D V_G$ at various $V_D$ for L=0.5μm, W=1μm and $t_{ox}$=2nm with interface trap distribution given in Fig. 4 at 4K.

It can be seen that the SS as a function of temperature still resembles the general observation in the experiment and the $I_D V_G$ curve shapes are similar at various $V_D$ (Fig. 7).

It is interesting to point out that, due to the abnormal SS, while there is severe degradation in SS due to DIBL at high temperatures (Fig. 6 and Fig. 8), the degradation is much less at cryogenic temperatures (Fig. 7 and Fig. 8) and this is expected to affect the device optimization strategy at cryogenic temperature.

## IV. CONCLUSIONS

To the best of our knowledge, for the first time, TCAD is used to model the *ON*-state characteristics and SS of nMOSFET using a single set of parameter and settings at 300K and 5K and verified against the experiment. In particular, it is showed that oxide/channel interface acceptor traps near the conduction band edge (mostly within 30meV) can be used to model the abnormal SS from 300K to 4K with reasonable density (about $10^{12}$cm$^{-2}$ for 2nm oxide). This validity is maintained under DIBL effect. This work does not prove or disprove any origin of the abnormal SS observed experimentally. However, it shows the possibility that the abnormal SS can be fully or partially due to the interface traps. Our setup provides a method to simulate and optimize cryogenic electronics in TCAD before the mechanism is fully understood.

ACKNOWLEDGMENT

This material is based upon work supported by the National Science Foundation under Grant No. 2046220.

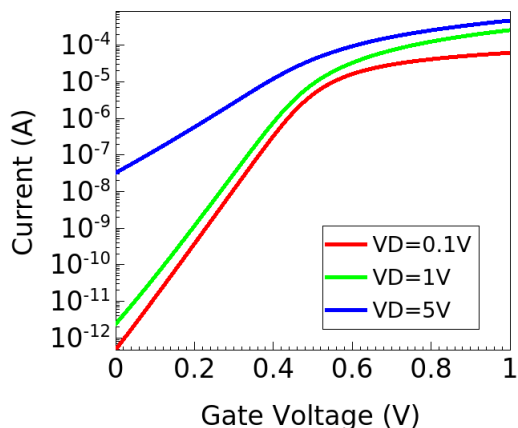

Figure 8: $I_D V_G$ at various $V_D$ for L=0.5μm, W=1μm and $t_{ox}$=2nm with interface trap distribution given in Fig. 4 at 300K.